\documentclass[amsmath,amsfonts,prd,nofootinbib,letterpaper,preprintnumbers,superscriptaddress]{revtex4}
\usepackage{graphicx}

\def\qslash{q\hspace{-1.5mm}/}

\def\Dslash{D\!\!\!\!/}
\def\ttwo{\mbox{$\frac{t}{2}$}}
\def\op{{\cal O}}

\begin{document}

\preprint{NT@UW-05-05, INT-PUB-05-16, UW/PT 05-17}

\title{Deep-inelastic scattering and the operator product expansion in
  lattice QCD }

\author{William Detmold} \affiliation{Department of Physics,
  University of Washington, Box 351560, Seattle, WA 98195, U.S.A.}

\author{C.-J. David Lin} \affiliation{Department of Physics,
  University of Washington, Box 351560, Seattle, WA 98195, U.S.A.}
\affiliation{Institute for Nuclear Theory, University of Washington,
  Box 351550, Seattle, WA 98195, U.S.A.}

\begin{abstract}
  We discuss the determination of deep-inelastic hadron structure in
  lattice QCD. By using a fictitious heavy quark, direct calculations
  of the Compton scattering tensor can be performed in Euclidean space
  that allow the extraction of the moments of structure functions.
  This overcomes issues of operator mixing and renormalisation that
  have so far prohibited lattice computations of higher moments.  This
  approach is especially suitable for the study of the twist-two
  contributions to isovector quark distributions, which is practical
  with current computing resources. While we focus on the isovector
  unpolarised distribution, our method is equally applicable to other
  quark distributions and to generalised parton distributions. By
  looking at matrix elements such as $\langle\pi^\pm| T [V^\mu(x)
  A^{\nu}(0)]|0\rangle$ (where $V^\mu$ and $A^\nu$ are vector and
  axial-vector heavy-light currents) within the same formalism,
  moments of meson distribution amplitudes can also be extracted.
\end{abstract}

\date\today \maketitle


\section{Introduction}

Lattice QCD offers the prospect of exploring the structure functions
probed in deeply inelastic scattering (DIS) and other high-energy
experiments from first principles.  By comparing to accurate
experimental data, such calculations provide stringent tests of QCD.
They also allow the extraction of information on hadron structure
which is not currently available from experiment, {\it e.g.}, the
transversity distribution $\delta q(x)$.  The structure functions
describe the hadronic part of the DIS process, {\it viz.}, the
hadronic tensor
\begin{equation}
\label{eq:W_def}
W^{\mu\nu}_{S}(p,q) = \int d^{4} x \hspace*{0.05cm}{\mathrm{e}}^{i
  q\cdot x} \langle p,S |\hspace*{0.03cm} \left [ J^{\mu}(x), J^{\nu}(0)
\right ] | p,S\rangle , 
\end{equation}
where $p$ and $S$ are the momentum and spin of the external state, $q$
is the momentum transfer between the lepton and the hadron, and
$J^{\mu}$ is the electromagnetic current.  Using the optical theorem,
$W^{\mu\nu}_{S}$ can be related to the imaginary part of the forward
Compton scattering tensor
\begin{equation}
\label{eq:T_def}
T^{\mu\nu}_{S}(p,q) = \int d^{4} x \hspace*{0.05cm}{\mathrm{e}}^{i
  q\cdot x} \langle p,S |\hspace*{0.03cm} T\left [ J^{\mu}(x)
  J^{\nu}(0)\right ] | p,S\rangle\, .  
\end{equation}
Since lattice QCD is necessarily formulated in Euclidean space, direct
calculation of the structure functions is challenging because of the
analytical continuation to Minkowski space that is required
\cite{Aglietti:1998mz, Phys.Rev.D62.074501}.  In addition, such a
calculation would involve all-to-all light-quark propagators, and is
therefore numerically demanding.  For these reasons, beginning with
pioneering works of Refs.~\cite{Kronfeld:1984zv,Martinelli:1987zd},
lattice studies of the deep-inelastic structure of hadrons have
focused on calculations of matrix elements of local operators that
arise from the light-cone operator product expansion (OPE)
\cite{Phys.Rev.179.1499,Christ:1972ms,Gross:1973ju,Gross:1974cs,Georgi:1951sr}
of the currents
\begin{equation}
  \label{eq:1}
  T[J^\mu(x) J^\nu(0)]=\sum_{i,n}  {\cal C}_i\left (x^2,\mu^2\right )
    \hspace*{0.05cm} x_{\mu_1}\ldots
  x_{\mu_n} \op^{\mu\nu\mu_1\ldots\mu_n}_i(\mu) ,
\end{equation}
where the ${\cal C}_i$ are the perturbatively calculable Wilson
coefficients that incorporate the short-distance physics, and the sum
is over all local operators, $\op_{i}^{\mu\nu\mu_1\ldots\mu_n}$ with
the correct symmetries. $\mu$ is the renormalisation scale. This
expansion enables the investigation of $T^{\mu\nu}_{S}$ via the
knowledge of hadronic matrix elements of local operators.  The
analytical continuation of these matrix elements from Euclidean space
to Minkowski space is straightforward.  However, a number of
difficulties arise in this approach because of the lattice regularisation.
Firstly, the non-zero lattice spacing breaks the symmetry group of
Euclidean space-time from $O(4)$ to the discrete hyper-cubic subgroup
$H(4)$, consequently modifying the transformation properties of the
local operators in the OPE. In general, operators belonging to
different irreducible representations of $O(4)$, which span the
right-hand side of the OPE in Eq.~(\ref{eq:1}), mix unavoidably in the
lattice theory since $H(4)$ has only a finite set of irreducible
representations.  For twist-two (twist = dimension - spin)
contributions, this becomes particularly severe for operators of spin
$n>4$ as they mix with lower dimensional operators and the mixing
coefficients contain power divergences.  Currently this restricts the
available lattice calculations to operators of spin $n=1,2,3,4$.  For
higher-twist operators, such power divergences are generally
unavoidable \cite{Gockeler:2000ja,Gockeler:2005vw,Gockeler:2001xw}.  A
second issue is that the matching of the lattice regularisation to
continuum renormalisation schemes
\cite{Martinelli:1994ty,Gockeler:1998ye,Capitani:2002mp,Guagnelli:2004ga},
in which the Wilson coefficients are calculated, becomes more involved
as $n$ increases.
In this paper, we discuss an approach to determining matrix elements
of higher-spin, twist-two operators in Eq.~(\ref{eq:1}).  This
approach is based upon directly studying the OPE on the lattice, as
was first investigated in kaon physics in Ref.~\cite{Dawson:1997ic}. A
similar technique has also been applied to determine Wilson
coefficients non-perturbatively \cite{Capitani:1998fe} and extract the
lowest moment of the isovector twist-two quark distribution
\cite{Capitani:1999fm} (our method is related to this latter work but
improves on it in a number of ways). In our proposal, one simulates
the Compton scattering tensor using lattice QCD, with currents
coupling the physical light quarks, $\psi (x)$, present in the hadron
to a non-dynamical (purely valence), unphysically heavy quark, $\Psi
(x)$.\footnote{Such fictitious currents have been used to study
  quark-hadron duality in heavy quark effective theory
  \cite{Shifman:1994yf}.  However, in this context the heavy quark is
  not a valence quark.}  The introduction of this heavy quark
significantly simplifies the calculation of isovector matrix elements
because it removes the requirement of all-to-all propagators.  After
performing an extrapolation to the continuum limit, the lattice data
for the Compton tensor are compared to the predictions of the OPE in
Euclidean space to extract the matrix elements of local operators in
Eq.~(\ref{eq:1}), directly in the continuum renormalisation scheme in
which the Wilson coefficients are calculated.  This approach also
removes the power divergences, thereby enabling extraction of matrix
elements of higher spin ($n > 4$) operators for twist-two operators
with a simple renormalisation procedure.  These matrix elements
determine the Mellin moments of the structure functions which are
identical in Euclidean space and Minkowski space and their analytical
continuation is trivial.  Finally, the chiral and infinite volume
extrapolations can now be performed at the level of the local matrix
elements using chiral perturbation theory
\cite{Detmold:2001jb,Arndt:2001ye,Chen:2001eg,Chen:2001gr,Beane:2002vq,
  Chen:2001yi,Detmold:2002nf,Detmold:2003tm,Detmold:2005pt}.
The matrix elements obtained via the above procedure are completely
independent of the mass of the unphysical, heavy quark and are indeed
physical quantities.  This is because such a quark can only propagate
between the bilocal currents, and the OPE relegates its short-distance
information to the Wilson coefficients.  In addition to the numerical
advantage, it also proves useful to introduce a fictitious heavy quark
for other reasons.  Firstly, the presence of the heavy scale
suppresses long distance correlations between the currents in a
similar way to a large Euclidean momentum.  Combining both the heavy
quark mass, $m_{\Psi}$, and momentum injection, $q$, at the current
allows us to control the behaviour of the OPE precisely at moderate
$m_{\Psi}$ and $q^{2}$.  The only constraint is
\begin{equation}
\label{eq:hierarchy}
\Lambda_{{\mathrm{QCD}}} \ll m_{\Psi} \sim \sqrt{q^{2}} \ll
\frac{1}{\hat{a}} , 
\end{equation}
where $\hat{a}$ is the coarsest lattice spacing used in the
calculation.  Secondly, the non-dynamical nature of the heavy quark
automatically removes many contributions (for example, so-called
``cat's ears'' diagrams -- see Fig.~\ref{fig:compton}(d) below) that are
higher-twist contaminations in traditional DIS.

In Section~\ref{sec:heavy}, we review the formalism of DIS with heavy
quarks before discussing the extraction of the moments of twist-two
parton distributions from lattice correlators in
Section~\ref{sec:lattice}. Finally in Section~\ref{sec:DAs}, we
broaden the analysis to investigate moments of meson distribution
amplitudes.


\section{Flavour changing currents and heavy quarks in lepton-hadron
  deep-inelastic scattering}
\label{sec:heavy}

The roles of quark and hadron masses in deep-inelastic scattering have
been well studied.  Target mass effects were first discussed by
Nachtmann \cite{Nachtmann:1973mr} and extensively investigated
throughout the 1970s, following the observation of the precocious
scaling of the structure functions \cite{Bloom:1970xb,Bloom:1971ye}.
Away from the Bjorken limit, they result in significant contributions
which arise from the OPE being an expansion in terms of operators
belonging to definite irreducible representations of the Lorentz
group.  These contributions scale as powers of $M^2/Q^2$, where $M$ is
the target mass and $Q^{2}=-q^{2}$, and can be summed exactly
\cite{Nachtmann:1973mr,Georgi:1976vf,Georgi:1976ve,Wandzura:1977ce}.
The effects of the struck and produced quark masses were also
comprehensively investigated
\cite{Witten:1975bh,Georgi:1976vf,Georgi:1976ve}.  These target and
quark mass effects lead to $\xi$ scaling
\cite{Nachtmann:1973mr,Georgi:1976vf,Georgi:1976ve,Wandzura:1977ce},
and are particularly relevant at moderate values of $Q^{2}$.  Since
currently available lattice cut-offs are $1/a \sim 3$ GeV, it is
important to include these mass effects in the application of the OPE
on the lattice, because of the condition in Eq.~(\ref{eq:hierarchy}).
In this section we present the OPE in Euclidean space relevant for
computing higher moments of parton distributions on the lattice with
these mass effects taken into account.
We consider fictitious currents that couple light up and down quarks
to unphysical heavy quarks of mass $m_{\Psi}$.  We focus on a purely
vector coupling, leaving the discussion of other possible currents to
the end of the section. We define
\begin{equation}
  \label{eq:2}
  J_{\Psi,\psi}^{\mu}(x)=\overline{\Psi}(x)\gamma^\mu \psi(x) +
  \overline{\psi}(x)\gamma^\mu \Psi(x)\,, 
\end{equation}
and construct the Euclidean Compton scattering tensor
\begin{equation}
  \label{eq:3}
  T^{\mu\nu}_{\Psi,\psi}(p,q)\equiv \sum_{S} \langle p, S|\hspace*{0.03cm}
  t^{\mu\nu}_{\Psi,\psi}(q)|p, S\rangle 
= \sum_{S} \int d^4x\ {\mathrm e}^{i q\cdot x} \langle p, S|
  T \left [ J_{\Psi,\psi}^{\mu}(x) J_{\Psi,\psi}^{\nu}(0)\right ] |p, S\rangle\,,
\end{equation}
(henceforth all momenta are Euclidean).
\begin{figure}[!t]
  \centering \includegraphics[width=0.3\columnwidth]{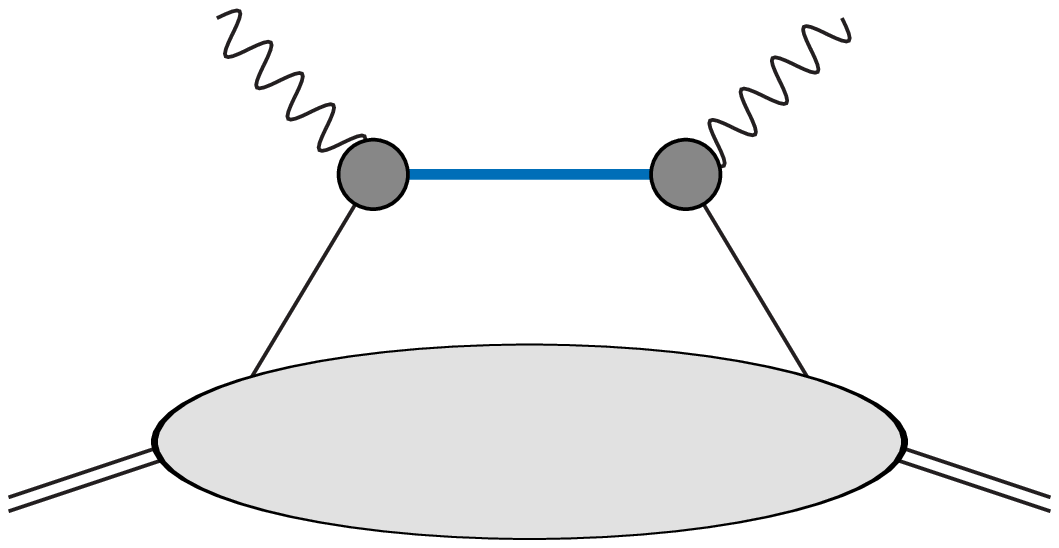}
  \includegraphics[width=0.3\columnwidth]{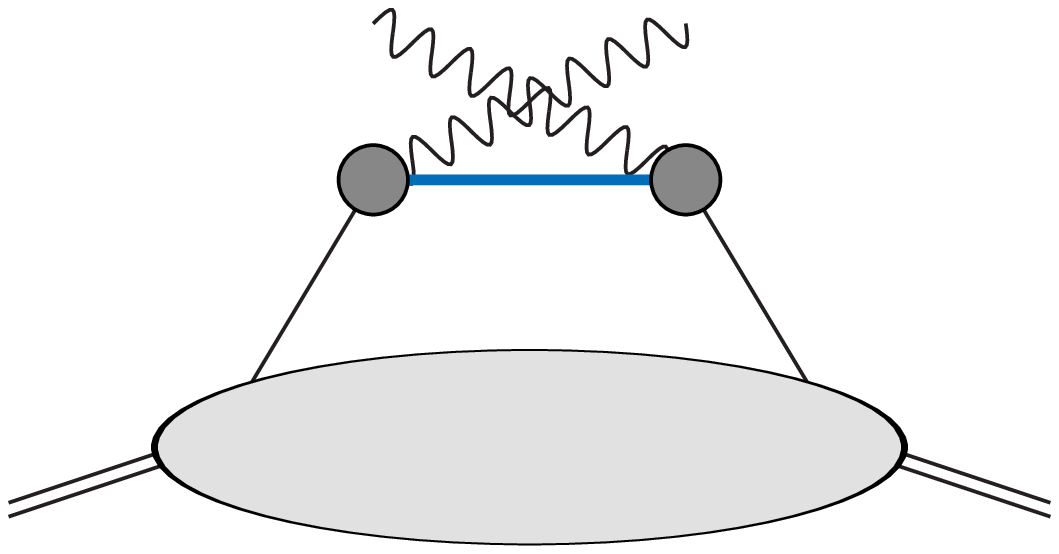}\\
  (a)\hspace*{0.3\columnwidth}(b)\\
  \includegraphics[width=0.3\columnwidth]{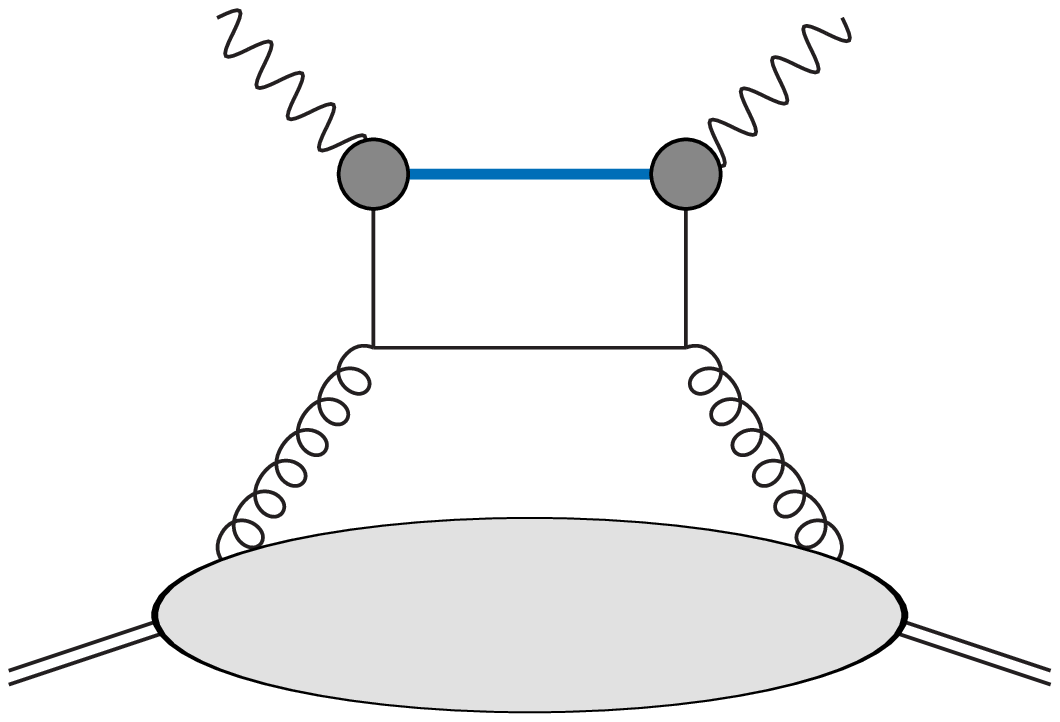}
  \includegraphics[width=0.3\columnwidth]{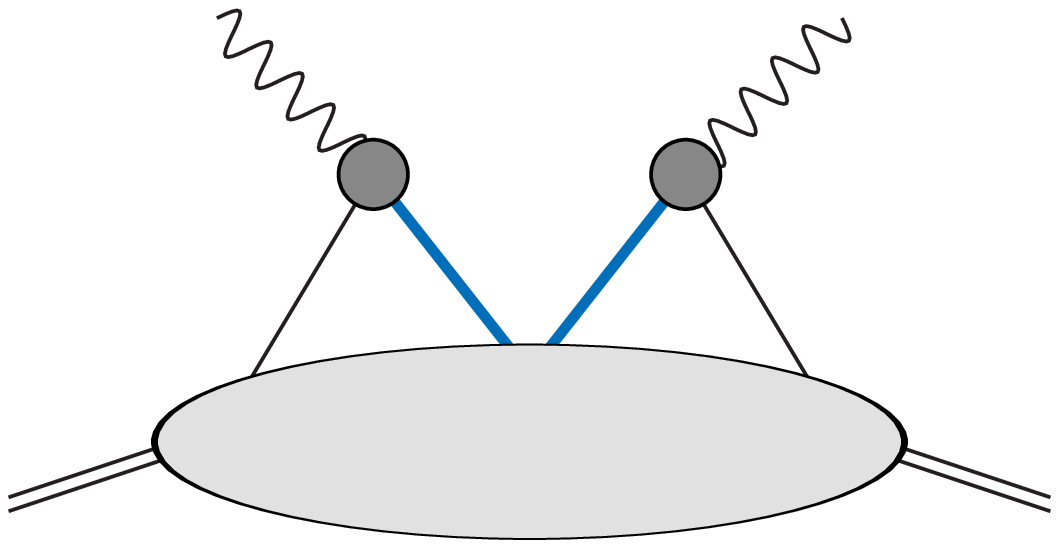}
  \includegraphics[width=0.3\columnwidth]{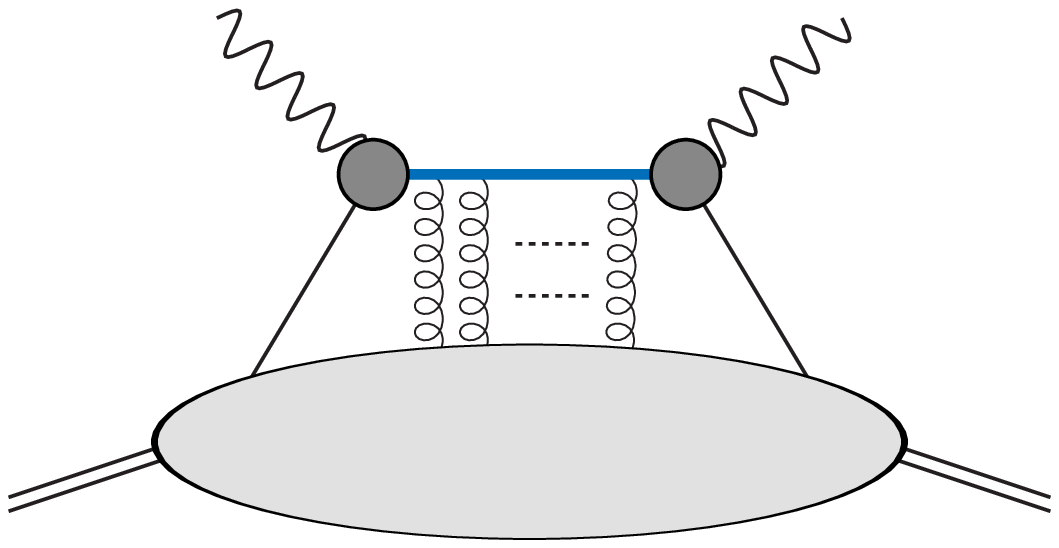}\\
  (c)\hspace*{0.285\columnwidth}(d)\hspace*{0.285\columnwidth}(e)
  \caption{Contributions to the Compton scattering tensor. Diagrams
    (a), (b) and (c) correspond to the leading twist contributions.
    Diagram (c) (the ``box diagram'') involves gluonic operators and
    vanishes for the isovector combination,
    Eq.~(\protect\ref{eq:isovector_tensor}). Diagram (d) (the ``cat's
    ears diagram'') is higher-twist and absent in our analysis.
    Diagram (e) includes leading- and higher-twist terms and is
    discussed in the main text. The thick lines correspond to the
    heavy-quark propagators, the shaded circles to the heavy-light
    currents and the large shaded regions to the various parton
    distributions.}
  \label{fig:compton}
\end{figure}
In the limit $q^{2} \rightarrow \infty$ or $m_{\Psi}\rightarrow
\infty$, $T^{\mu\nu}_{\Psi,\psi}$ is given by the leading-twist
contribution, the ``handbag diagrams'' in Figs.~\ref{fig:compton} (a)
and (b). The ``box diagram'', Fig.~\ref{fig:compton} (c)\footnote{In
  Fig.~\ref{fig:compton} (c), we specify that the large momentum,
  $q^{2}$, flows through the three light-quark lines; the
  contributions in which these quarks have soft momenta are already
  included in Figs.~\ref{fig:compton} (a) and (b).  In principle,
  these gluonic contributions can be disentangled from their different
  $q^{2}$ behaviour.}, which involves purely gluonic operators after
the OPE, is strongly suppressed in our approach and is completely
absent in the study of the OPE of the isovector Compton scattering
tensor
\begin{equation}
\label{eq:isovector_tensor}
T^{\mu\nu}_{\Psi,v} = T^{\mu\nu}_{\Psi,u} - T^{\mu\nu}_{\Psi,d} .
\end{equation}
This makes the extraction of moments of the isovector quark
distributions practical, and we focus on this case in this paper.
At moderate $q^{2}$ and $m_\Psi$, higher-twist terms also contribute.
However, the non-dynamical nature of the fictitious heavy quark
entirely eliminates the higher-twist contributions involving more than
one quark propagator between the currents, {\it e.g.}, the ``cat's
ears diagram'' in Fig.~\ref{fig:compton} (d).  The diagrams in
Fig.~\ref{fig:compton} (e) contain pieces that contribute to the
twist-two operators in Eqs.~(\ref{eq:7}) and (\ref{eq:8}), and
also higher-twist terms that are discussed below.
The twist-two contributions to the OPE in $T^{\mu\nu}_{\Psi,v}$ are
from
\begin{equation}
  \label{eq:4}
  t_{\Psi,\psi}^{\mu\nu}=\overline{\psi}\gamma^\mu
  \frac{-i\left(i\tensor{\Dslash}+\qslash\right)+m_{\Psi}}
    {(i\tensor{D}+q)^2+m_{\Psi}^2}  \gamma^\nu \psi\,,
\end{equation}
and a similar term, Fig.~\ref{fig:compton} (b), in which
$\mu\leftrightarrow\nu$ and $q\to-q$. The derivatives
[$\tensor{D}^\mu=\frac{1}{2}\left(\roarrow{D}^\mu-\loarrow{D}^\mu\right)$]
are included to account for the soft transverse momentum of the struck
quark; they are covariant in order to maintain gauge-invariance.
The OPE of the bilocal currents in Eq.~(\ref{eq:3}) is now given by a
Taylor expansion of the propagator in the above expression,
\begin{equation}
  \label{eq:5}
    \frac{-i\left(i\tensor{\Dslash}+\qslash\right)+m_{\Psi}}
    {(i\tensor{D}+q)^2 + m_{\Psi}^{2}}  =
  - \frac{-i\left(i\tensor{\Dslash}+\qslash\right)+m_{\Psi}}
    {Q^2+\tensor{D}^2 - m_{\Psi}^{2}}  \sum_{n=0}^{\infty}\left(\frac{-2i\ q\cdot
  \tensor{D}}{Q^2+\tensor{D}^2-m_{\Psi}^{2}}\right)^n \,,
\end{equation}
where $Q^2=-q^2$. The pole mass of the heavy quark is not a measurable
quantity and we replace it with the mass of a meson composed of the
unphysical heavy quark and a light quark,
$M_{\Psi}=m_{\Psi}+\frac{1}{2}\alpha$, where $\alpha$ is the binding
energy and is of $\op(\Lambda_{{\mathrm{QCD}}})$.  This meson mass can
be directly computed on the lattice.  The parameter $\alpha$ can be
extracted from experiment or lattice calculations
\cite{El-Khadra:2002wp}, however we view it as an unknown to be
determined in the procedure of studying the OPE on the lattice.  The
$\tensor{D}^{2}$ term in the denominators in Eq.~(\ref{eq:5}) gives
rise to higher-twist contributions, such as those arising from
Fig.~\ref{fig:compton} (e), if we instead Taylor expand with respect
to $\left(\frac{-2i q\cdot \tensor{D} +
    \tensor{D}^{2}}{Q^2-m_{\Psi}^{2}}\right)$.  These higher-twist
contributions scale as powers of $\Lambda_{{\mathrm{QCD}}}^{2}/(q^{2}
+ m_{\Psi}^{2})$ and can be neglected for $q^{2}$ or $m_{\Psi}^{2}$
sufficiently large.  For more moderate scales, they may become
important (particularly for large $n$) and can in principle be studied
in our approach.  However, this is beyond the scope of this work.
Instead, we replace the denominator in Eq.~(\ref{eq:5}) by
\begin{equation}
\label{eq:q_tilde}
\widetilde{Q}^2=Q^2-M_{\Psi}^{2}+\alpha M_{\Psi} + \beta, 
\end{equation}
where the unknown parameter $\beta$ represents terms of
$\op(\Lambda_{{\mathrm{QCD}}}^{2})$, including these higher-twist
effects and sub-leading heavy-quark mass effects. In doing so we have
neglected the $n$ dependence of $\beta$ since it is suppressed by
powers of the strong coupling. To remove these higher-twist and mass
uncertainties, one might consider using a fictitious heavy particle
that does not interact strongly. However, issues such as the gauge
dependence of the resulting Compton tensor would need to be
investigated thoroughly and are not discussed here.
After making this expansion and considering only the symmetric
combination ($m$ is the light quark mass)
$t^{\{\mu\nu\}}_{\Psi,\psi}=\frac{1}{2}
\left(t^{\mu\nu}_{\Psi,\psi}+t^{\nu\mu}_{\Psi,\psi}\right)$, combining
the two leading contributions then leads to
\begin{eqnarray}
  t_{\Psi,\psi}^{\{\mu\nu\}}(q)&=& \frac{i}{\widetilde{Q}^2}
  \Bigg(4 \sum_{\substack{n=0 \\ {\rm even}}}^\infty {\cal C}_{n+2}\frac{(2
    q_{\mu_1}) \ldots (2 
      q_{\mu_{n}})}{\widetilde{Q}^{2n}} {\cal
        O}_{\psi}^{\mu\nu\mu_1\ldots\mu_{n}}  
    + \widetilde{Q}^2 \delta^{\mu\nu}\sum_{\substack{n=2 \\ {\rm
          even}}}^\infty  {\cal C}_{n}^\prime \frac{(2
      q_{\mu_1}) \ldots (2 
      q_{\mu_{n}})}{\widetilde{Q}^{2n}} {\cal
        O}_{\psi}^{\mu_1\ldots\mu_{n}}  
  \label{eq:6}
\\ 
&& \hspace*{10mm}
+ 2 i \delta^{\mu\nu} (m_{\Psi}-m) \sum_{\substack{n=0 \\ {\rm even}}}^\infty
 \widehat{\cal C}_{n}\frac{(2 q_{\mu_1}) \ldots (2 
      q_{\mu_{n}})}{\widetilde{Q}^{2n}} \widehat {\cal
        O}_{\psi}^{\mu_1\ldots\mu_{n}}
- 4 q^{\{\mu} \sum_{\substack{n=1 \\ {\rm odd}}}^\infty  {\cal
  C}_{n+1}^{\prime\prime} \frac{(2
  q_{\mu_1}) \ldots (2 
      q_{\mu_{n}})}{\widetilde{Q}^{2n}} {\cal
        O}_{\psi}^{\nu\}\mu_1\ldots\mu_{n}}
\Bigg)    \,,
\nonumber 
\end{eqnarray}
where we have introduced the operators (the braces indicate
symmetrisation of the enclosed indices)
\begin{equation}
  \label{eq:7}
  \op_{\psi}^{\mu_1\ldots\mu_n}=\overline{\psi} \gamma^{\{\mu_1}\left(i
    \tensor{D}^{\mu_2}\right) \ldots \left(i
    \tensor{D}^{\mu_n\}}\right) \psi - {\rm traces} \,,
\end{equation}
and
\begin{equation}
  \label{eq:8}
  \widehat\op_\psi^{\mu_1\ldots\mu_n}=\overline{\psi} \left(i
    \tensor{D}^{\{\mu_1}\right) \ldots \left(i \tensor{D}^{\mu_n\}}\right)
  \psi - {\rm traces}\,, 
\end{equation}
and the various perturbatively calculable Wilson coefficients ${\cal
  C}_n$, ${\cal C}_n^\prime$, ${\cal C}_n^{\prime\prime}$ and
$\widehat{\cal C}_n$ depend on $q^2$, $\mu^2$ (the renormalisation
scale) and the heavy quark mass
\cite{Gottschalk:1980rv,Kramer:1991uh,Aivazis:1993kh,Aivazis:1993pi}.
The first set of operators are the usual twist-two vector operators
that enter into textbook DIS analyses.  The second set are chiral-odd,
twist-three operators whose matrix elements correspond to the parton
distribution $e_{\psi}(x)$ \cite{Jaffe:1991kp,Jaffe:1991ra}.
Taking spin-averaged hadron matrix elements of this expression then
leads to
\begin{eqnarray}
  \label{eq:9}
  T^{\{\mu\nu\}}_{\Psi,\psi}(p,q) &=& 
i  \sum_{\substack{n=2 \\ {\rm even}}}^\infty  A_{\psi}^n(\mu^2)
  \zeta^n \Bigg\{ \delta^{\mu\nu}
  \left[{\cal C}_{n}\frac{\widetilde{Q}^2}{q^2}\frac{n C_{n}^{(1)}(\eta) - 2\eta
      C_{n-1}^{(2)}(\eta)}{n(n-1)} + {\cal C}^\prime_{n} C_{n}^{(1)}(\eta)\right] 
+ \frac{p^\mu p^\nu \widetilde{Q}^2}{(p\cdot q)^2}{\cal C}_{n}\left[\frac{8\eta^2
    C_{n-2}^{(3)}(\eta)}{n(n-1)}\right] 
\\
&&
+4\frac{p^{\{\mu}q^{\nu\}}}{p\cdot q}
\left[
  {\cal C}_{n} \frac{\widetilde{Q}^2}{q^2}\frac{(n-1)\eta C_{n-1}^{(2)}(\eta)-4\eta^2
    C_{n-2}^{(3)}(\eta)}{n(n-1)} -{\cal C}_{n}^{\prime\prime}
  \frac{\eta}{n}C_{n-1}^{(2)}(\eta)\right] 
\nonumber \\
&&
+ \frac{q^\mu q^\nu}{q^2}
\left[
  {\cal C}_{n}\frac{\widetilde{Q}^2}{q^2}\frac{n(n-2)C_{n}^{(1)}(\eta)-2\eta(2n-3)
    C_{n-1}^{(2)}(\eta)+8\eta^2 C_{n-2}^{(3)}(\eta)}{n(n-1)}
  -2{\cal C}_{n}^{\prime\prime} \left( C_{n}^{(1)}(\eta)
    -2\frac{\eta}{n} C_{n-1}^{(2)}(\eta)\right)\right]  
\Bigg\}
\nonumber  \\
&&
-2i\frac{M(m_{\Psi}-m)}{\widetilde{Q}^2}\delta^{\mu\nu}  
\sum_{\substack{n=0 \\ {\rm even}}}^\infty \widehat{\cal C}_{n}
\widehat{A}_{\psi}^n(\mu^2)   \zeta^n C_{n}^{(1)}(\eta)\,,
\nonumber
\end{eqnarray}
where we have defined
\begin{equation}
\zeta=\frac{\sqrt{p^2q^2}}{\widetilde{Q}^2},
\quad\quad\eta=\frac{p\cdot q}{\sqrt{p^2q^2}}\,,
\end{equation}
($M$ is the proton mass, $p^2=-M^2$) and the hadronic matrix elements
of the local operators in Eqs.~(\ref{eq:7}) and (\ref{eq:8}) as
\begin{eqnarray}
  \label{eq:10}
  \sum_S\langle p,S | \op_{\psi}^{\mu_1\ldots\mu_{n}} | p,S \rangle 
  &=& A^n_{\psi}(\mu^2) \left[p^{\mu_1}\ldots p^{\mu_n}  - {\rm traces}\right]\,,
\\
  \label{eq:11}
  \sum_S\langle p,S | \widehat\op_{\psi}^{\mu_1\ldots\mu_{n}} | p,S
  \rangle  
&=& i\ M\ \widehat A^n_{\psi}(\mu^2) \left[p^{\mu_1}\ldots p^{\mu_n} - {\rm
    traces} \right]\,,
\end{eqnarray}
(the $A_{\psi}^n$ and $\widehat A_{\psi}^n$ are dimensionless and
real). In Eq.~(\ref{eq:9}) the $C_n^{(\lambda)}(\eta)$ are Gegenbauer
polynomials that arise from the trace subtractions in
Eqs.~(\ref{eq:10}) and (\ref{eq:11}) and account for the target mass
effects
\cite{Nachtmann:1973mr,Georgi:1976vf,Georgi:1976ve,Wandzura:1977ce}.
If we now choose the rest frame of the proton, $p=(0,0,0,i\;M)$ and
parameterise $q=(0,0,\sqrt{q_{0}^{2}-Q^2},i\;q_{0})$, then the symmetric
combination of $\{\mu,\nu\}=\{3, 4\}$ is
\begin{eqnarray}
  \label{eq:12}
  T_{\Psi,\psi}^{\{34\}}(p,q)&=&  
\sum_{\substack{n=2 \\ {\rm even}}}^{\infty} A_{\psi}^{n}(\mu^2) f(n)\,,
\end{eqnarray}
where
\begin{eqnarray}
\label{eq:13}
f(n)&=&-\sqrt{q_{0}^{2}-Q^2}\zeta^n\Bigg\{\frac{2}{q_{0}}
\Bigg[{\cal C}_{n}\frac{\widetilde{Q}^2}{Q^2}\frac{(n-1)\eta
  C_{n-1}^{(2)}(\eta)-4\eta^2 
  C_{n-2}^{(3)}(\eta)}{n(n-1)} +{\cal C}_{n}^{\prime\prime}
\frac{\eta}{n}C_{n-1}^{(2)}(\eta)
    \Bigg]
\\&&\hspace*{5mm}
+\frac{q_{0}}{Q^2}\Bigg[
{\cal C}_{n}\frac{\widetilde{Q}^2}{Q^2}\frac{n(n-2)C_n^{(1)}(\eta)
  -2\eta(2n-3)C_{n-1}^{(2)}(\eta) +8\eta^2
  C_{n-2}^{(3)}(\eta)}{n(n-1)}+
2{\cal C}_{n}^{\prime\prime}\left(C_{n}^{(1)}(\eta)-2\frac{\eta}{n}
  C_{n-1}^{(2)}(\eta)\right) 
\Bigg]\Bigg\}
 \,.
\nonumber 
\end{eqnarray}
Equation~(\ref{eq:12}) is the central object studied in the following
section where we will see that lattice calculations of it would allow
extraction of the even moments $A_{\psi}^n(\mu^2)$, since the Wilson
coefficients and kinematic factors that determine $f(n)$ are known.
However we discuss some more general aspects of
$T^{\mu\nu}_{\Psi,\psi}$ first.
Once we have determined the $A_{\psi}^n(\mu^2)$, we can also use the
diagonal elements of $T^{\mu\nu}_{\Psi,\psi}$ to extract information
on the moments of the twist-three quark distributions $e_{\psi}(x)$
from the same vector-vector current correlator.  Experimentally, this
distribution is difficult to extract, being measurable only in charged
current DIS, semi-inclusive DIS or Drell-Yan processes
\cite{Jaffe:1991kp,Jaffe:1991ra}. The only currently available
determination uses data from the CLAS collaboration
\cite{Avakian:2003pk} and requires assumptions about the fragmentation
functions that enter semi-inclusive DIS \cite{Efremov:2002ut}. Any
information on the moments of $e_{\psi}(x)$ from lattice QCD (whether
in our approach or by direct calculation) would be useful.
Additionally, the odd-$n$ moments $A_{\psi}^n(\mu^2)$ and
$\widehat{A}_{\psi}^n(\mu^2)$ (which determine the valence
combinations of quark distributions) can be extracted from a
correlator of vector and axial-vector currents just as for $F_3(x)$ in
neutrino scattering.
Similar methods can be used to calculate moments of the helicity and
transversity parton distributions. The moments of both the twist-two
and twist-three helicity distributions and the twist-two transversity
distribution may be determined from suitable antisymmetric pieces of
$T^{\mu\nu}_{\Psi,\psi}$. Since we are not restricted by physical
scattering processes, we can also consider correlators of unphysical
currents: {\it e.g.} scalar, pseudo-scalar and tensor. The analysis
for such currents is again very similar to the case that we have
discussed and by using such currents we can investigate distributions
that are not experimentally accessible in DIS; for example the
transversity distribution can be accessed independent of the quark
masses by looking at the scalar--vector correlator. Since these
currents are not conserved, they are no longer scale independent and
the appropriate anomalous dimensions must be used
\cite{Blumlein:2001ca,Ioffe:1994aa}.  Finally, the off-forward matrix
elements of the various twist-two and twist-three operators that
determine the moments of generalised parton distributions can also be
studied through analysis of the off-forward Compton scattering
amplitude.
%


\section{Extraction of moments from lattice calculations}
\label{sec:lattice}

\begin{figure}[!t]
  \centering \includegraphics[width=0.3\columnwidth]{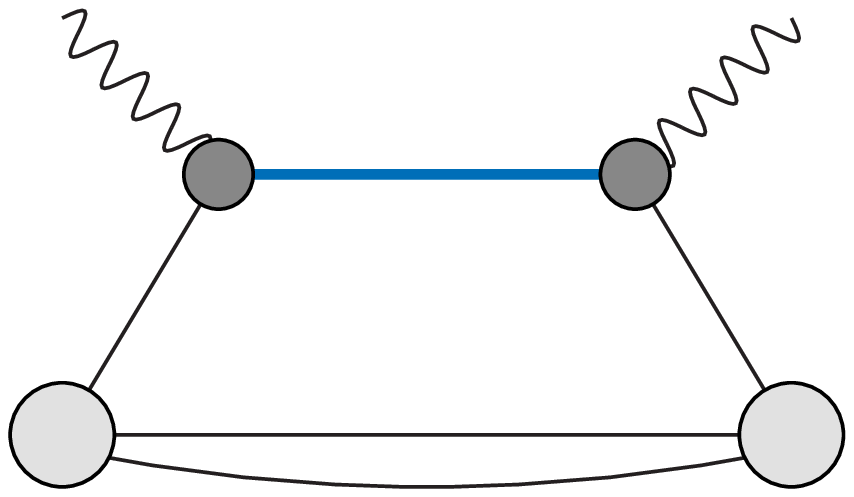}
  \hspace*{5mm} \includegraphics[width=0.3\columnwidth]{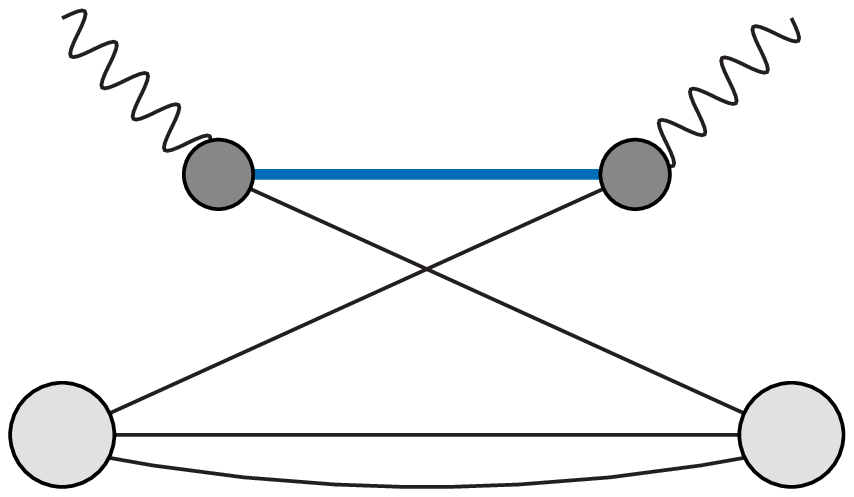}
  \hspace*{5mm}
  \includegraphics[width=0.3\columnwidth]{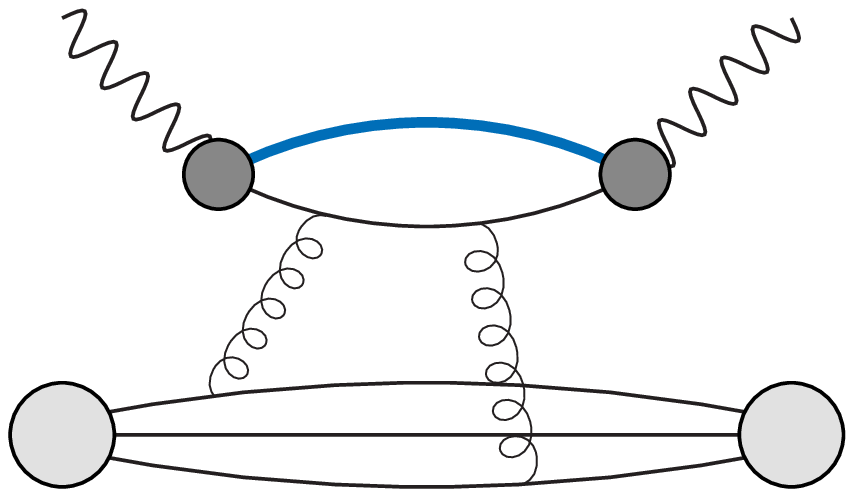} \\
  \centering (a) \hspace*{5.5cm} (b) \hspace*{5.5cm} (c)
  \caption{Quark contractions in the four point correlator. Light
    shaded circles are the proton source and sink whilst the dark
    shaded circles are the heavy-light currents. The thick line
    indicates the heavy quark propagator. Diagram (c) is quark-line
    disconnected (as indicated by the representative gluons).}
  \label{fig:quark_contractions}
\end{figure}

\begin{figure}[!t]
  \centering \includegraphics[width=0.3\columnwidth]{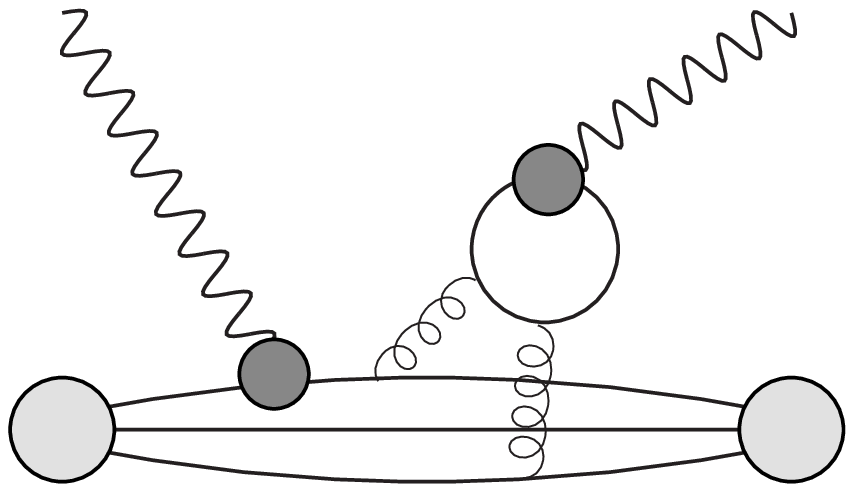}
  \hspace*{5mm} \includegraphics[width=0.3\columnwidth]{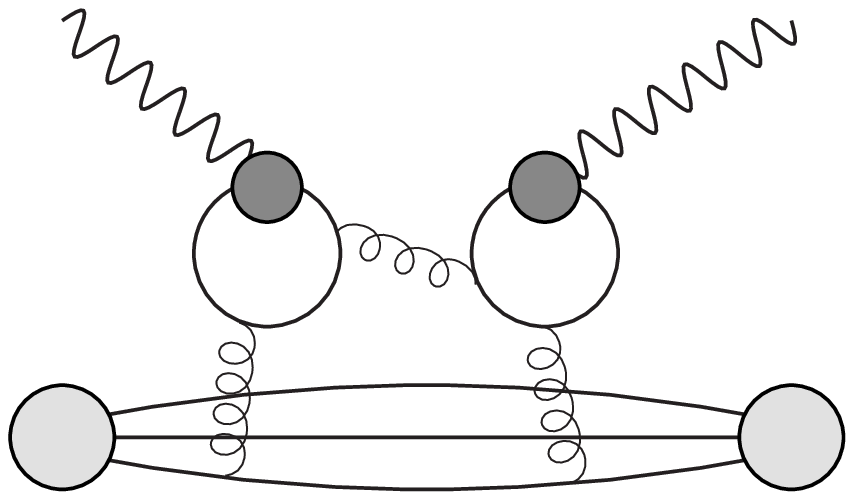}
  \hspace*{5mm}
  \includegraphics[width=0.3\columnwidth]{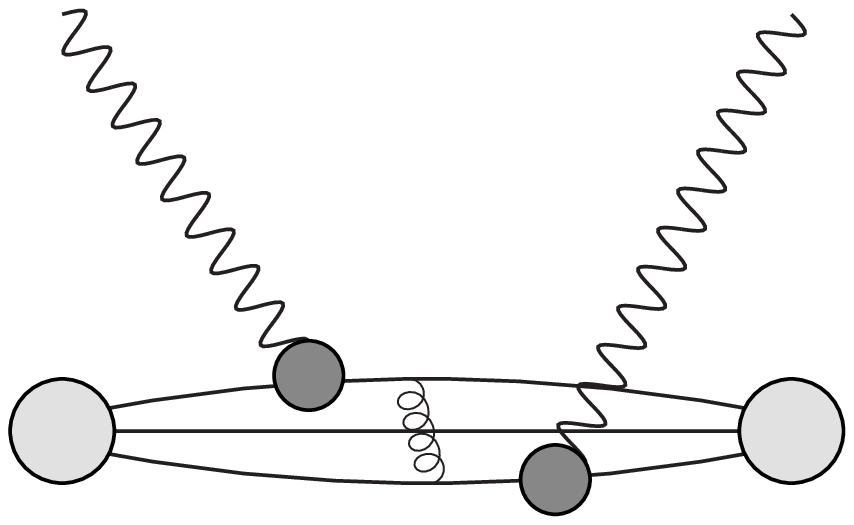} \\
  \centering (a) \hspace*{5.5cm} (b) \hspace*{5.5cm} (c)
  \caption{Additional quark contractions present in the four point correlator
    for light-light currents but absent in the case we consider. These
    diagrams correspond to higher-twist contributions (diagram (d) in
    Fig.~\protect{\ref{fig:compton}}). Light shaded circles are the
    proton source and sink whilst the dark shaded circles are the
    light-light currents.}
  \label{fig:missing_quark_contractions}
\end{figure}

\begin{figure}[!t]
  \centering
  \includegraphics[width=0.48\columnwidth]{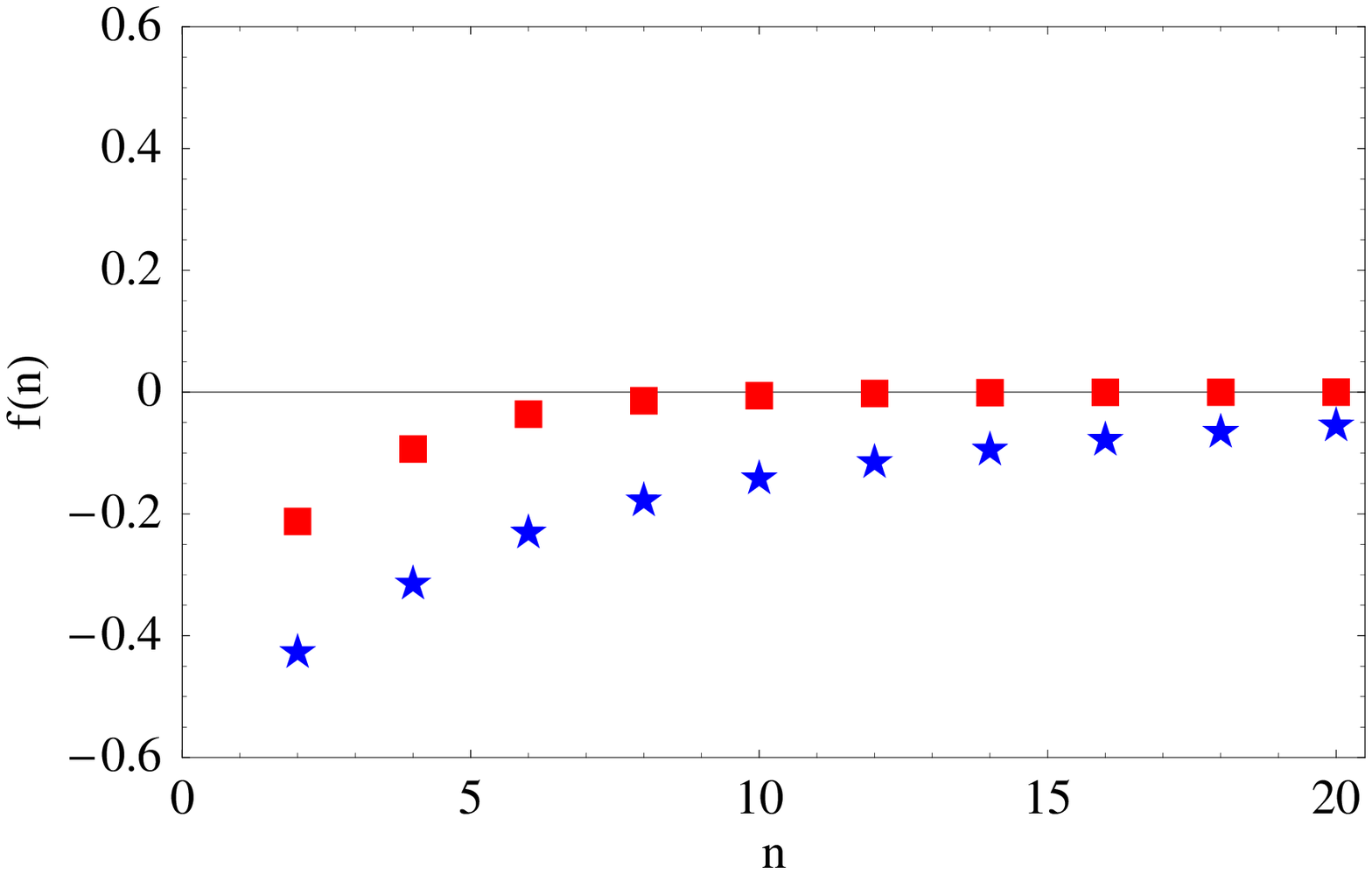}\hspace{5mm}
  \includegraphics[width=0.48\columnwidth]{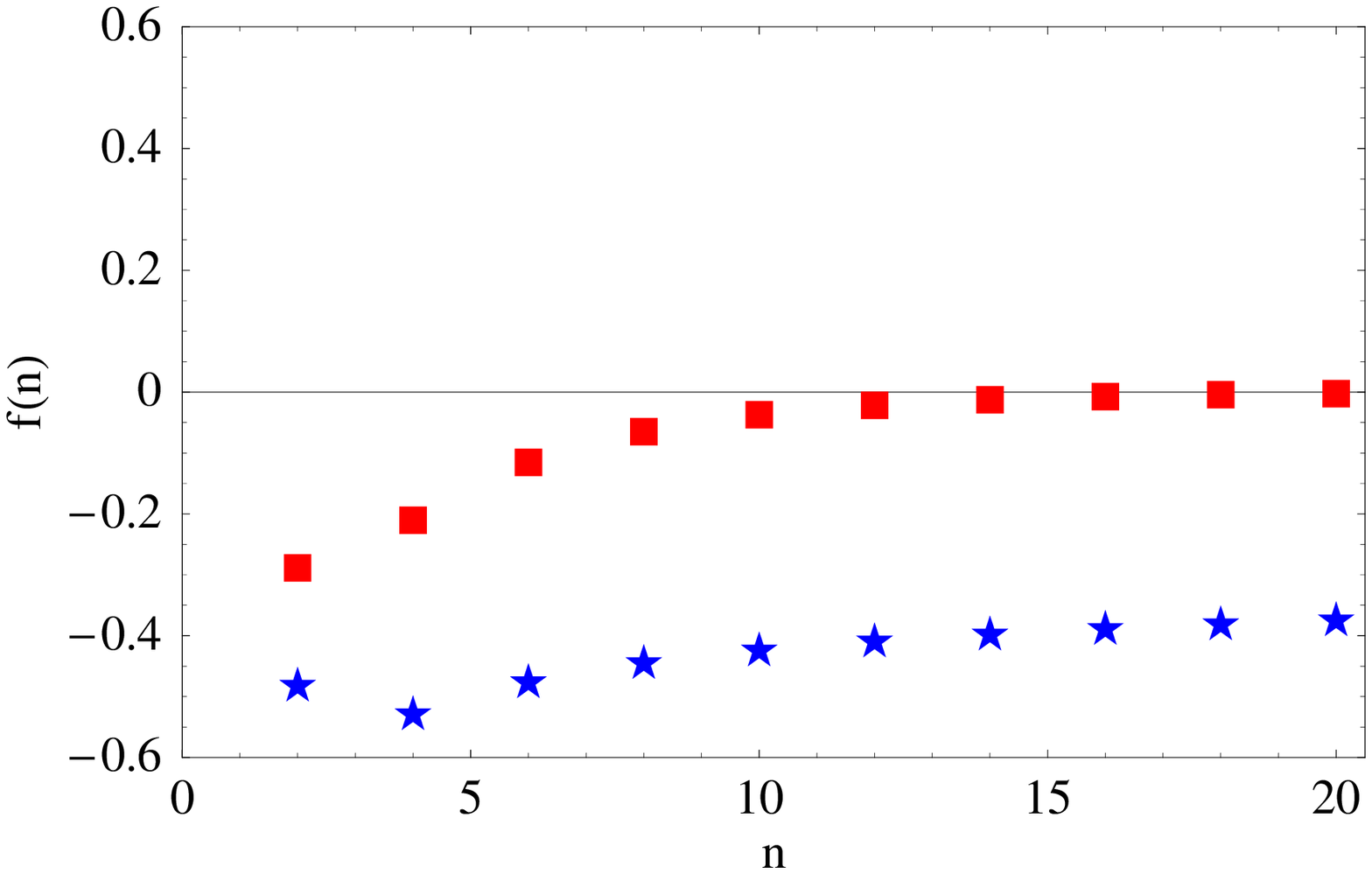}
  \caption{Behaviour of the function $f(n)$ for two different
    parameter sets. On the left, $M_{\Psi}=3.54$~GeV,
    $Q^2=1.5$~GeV$^2$ and $q_{0}=2.76$~GeV while on the right we have
    chosen $M_{\Psi}=2.1$~GeV, $Q^2=-3.85$~GeV$^2$ and
    $q_{0}=1.98$~GeV. The stars and boxes correspond to
    $\alpha=0.4,\,1.2$~GeV respectively and we have chosen $M=1.2$~GeV
    and $\beta=0$~GeV$^2$. }
  \label{fig:fn}
\end{figure}

With numerical investigation of $T^{\{34\}}_{\Psi,\psi}(p,q)$ for
varying $Q^2$, $q_{0}$ and $M_{\Psi}$, Equation~(\ref{eq:12}) provides a
means to extract the moments $A^{n}_{\psi}(\mu)$ for $n>4$ without
requiring power subtractions and complicated renormalisations. In
order to calculate $T^{\{\mu\nu\}}_{\Psi,\psi}(p,q)$ we consider the
following four-point Euclidean correlator
\cite{Phys.Rev.D62.074501,Liu:1993cv} (with $\tau>0$)
\begin{eqnarray}
  \label{eq:14}
G_{(4)}^{\mu\nu}({\bf p},{\bf q},t,\tau;\Gamma) &=& \sum_{{\bf x},{\bf z}}\sum_{\bf
  y} {\mathrm{e}}^{i {\bf
        p}\cdot{\bf x}}{\mathrm{e}}^{i {\bf q} \cdot {\bf y}} \Gamma_{\beta\alpha}
\langle 0|\chi_\alpha({\bf x},t))\overline{J}_{\Psi,\psi}^\mu({\bf
  y}+{\bf z},\tau + \ttwo)
\overline{J}_{\Psi,\psi}^\nu({\bf z},\ttwo)\overline{\chi}_\beta({\bf 0},0)|0\rangle 
\\
&=& \sum_{{\bf x},{\bf z}}\sum_{\bf y} \sum_{N,N^\prime}\sum_{s,s^\prime}
{\mathrm{e}}^{i ({\bf p}-{\bf p_N})\cdot{\bf x}}{\mathrm{e}}^{i ({\bf 
      p_N}-{\bf p_{N^\prime}})\cdot{\bf z}} {\mathrm{e}}^{i {\bf q} \cdot {\bf y} }
  {\mathrm{e}}^{-(E_N+E_{N^\prime})\ttwo} \Gamma_{\beta\alpha} 
\nonumber 
\\&& \hspace*{0.8cm}\times
\langle 0|\chi_\alpha(0)|E_N,{\bf p_N},s\rangle \langle E_N,{\bf
  p_N},s |\overline{J}_{\Psi,\psi}^\mu({\bf y},\tau)
\overline{J}_{\Psi,\psi}^\nu(0)|E_{N^\prime},{\bf p_{N^\prime}},s^\prime\rangle
\langle E_{N^\prime},{\bf p_{N^\prime}},s^\prime
|\overline{\chi}_\beta(0)|0\rangle 
\nonumber \\
&\hspace*{-2mm}\stackrel{t\rightarrow\infty}{\longrightarrow}&
{\mathrm{e}}^{-E_0 t}\sum_{{\bf y}} {\mathrm{e}}^{i{\bf q}\cdot
  {\bf y}}\sum_{s,s^\prime}\Gamma_{\beta\alpha} 
\langle 0|\chi_\alpha(0)|E_0,{\bf p},s\rangle \langle E_0,{\bf
  p},s |\overline{J}_{\Psi,\psi}^\mu({\bf y},\tau)
\overline{J}_{\Psi,\psi}^\nu(0)|E_{0},{\bf p},s^\prime\rangle
\langle E_0,{\bf p},s^\prime
|\overline{\chi}_\beta(0)|0\rangle  \,,
\nonumber
\end{eqnarray}
where $\chi$ is a dimensionless interpolating field for the proton and
$\overline{J}^\mu_{\Psi,\psi}$ is the lattice version of
Eq.~(\ref{eq:2}). Here $\Gamma$ is a Dirac matrix which can be chosen
as $\Gamma_4=\frac{1}{2}(1+\gamma_4)$ for the components of the
amplitude we are considering and $E_0({\bf p})$ is the ground state
energy.  The sums on $N$ and $N^{\prime}$ are over all possible
eigenstates of the Hamiltonian.  Then, defining the two-point
Euclidean correlator as
\begin{equation}
  \label{eq:15}
  G_{(2)}({\bf p},t;\Gamma)=\sum_{\bf x} {\mathrm e}^{i {\bf p}\cdot{\bf
      x}}\Gamma_{\beta\alpha} 
\langle0|\chi_\alpha(0)\overline{\chi}_\beta({\bf x},t)|0\rangle ,
\end{equation}
we can determine the Compton amplitude from the Fourier transform of
the ratio of these correlators:
\begin{equation}
  \label{eq:16}
T^{\{\mu\nu\}}_{\Psi,\psi}(p,q)=4M\, a \sum_\tau {\mathrm
  e}^{iq_4\tau}\left[\lim_{t\to\infty} 
  \frac{ G^{\{\mu\nu\}}_{(4)}({\bf
    p},{\bf q},t,\tau;\Gamma_4)}{G_{(2)}({\bf p},t;\Gamma_4)}\right]\,,
\end{equation}
where $a$ is the lattice spacing. Since $G_{(4)}^{\mu\nu}({\bf p},{\bf
  q},t,\tau;\Gamma)$ falls off with $\tau$ over a characteristic time
$(M_\Psi a)^{-1}$, the Fourier transform in Eq.~(\ref{eq:16}) will be
well approximated in practice.
The above equations also hold if the heavy-light current is replaced
by the usual light-light current. However, using a quenched heavy
quark greatly simplifies the numerical work in extracting
$T^{\{34\}}_{\Psi,\psi}(p,q)$ without altering the non-perturbative
physics, the Mellin moments, we are interested in.  After performing
the Wick contractions of the quark fields, $G_{(4)}^{\mu\nu}({\bf
  p},{\bf q},t,\tau;\Gamma)$ is given by the three different
arrangements of quark propagators shown in
Fig.~\ref{fig:quark_contractions} where the thick and thin lines
represent the heavy and light (physical) quarks respectively. For the
isovector channel obtained via Eq.~(\ref{eq:isovector_tensor}),
diagram \ref{fig:quark_contractions} (c) does not contribute. In the
remaining two diagrams, the Wick contractions can be computed with the
technique of extended propagators. Many values of the heavy quark mass
can be studied with only a very small increase in computational cost.
For a light-light current, we would need the additional Wick
contractions shown in Fig.~\ref{fig:missing_quark_contractions} and
would require light all-to-all propagators even in the isovector
channel.
In order to avoid the problems of operator mixing discussed in the
introduction, a continuum extrapolation of the Compton amplitude must
be performed before it can be used to extract the moments
$A_{\psi}^n(\mu)$ in Eq.~(\ref{eq:12}).  This requires calculations at
a number of different lattice spacings. On the other hand, the chiral
and infinite volume extrapolations can only be performed after we have
extracted the local matrix elements since the current-current
correlator is not a low-energy observable and cannot be treated
reliably in effective field theory. Also, to use Eq.~(\ref{eq:12}), the
heavy-light meson and proton masses must be extracted from appropriate
two-point correlators.
Assuming these tasks have been performed, we may now consider what is
required to obtain useful information from these calculations.
Ideally we would extract all integer moments from the lattice and be
able to uniquely determine the Minkowski space parton distributions
through an inverse Mellin transform.\footnote{This is guaranteed to be
  a unique reconstruction by Carlson's theorem \cite{CarlsonThm}.
  Essentially we would be expanding the OPE in the Euclidean region,
  analytically continuing the Wilson coefficients and then re-summing
  the expansion in the Minkowski region.} More realistically, we might
envisage extracting 6--10 even moments (and the parameters $\alpha$ and
$\beta$ in Eq.~(\ref{eq:q_tilde})) by fits using Eq.~(\ref{eq:12}).
Previous analysis \cite{Detmold:2001dv} suggests that this is enough
to reliably constrain standard parameterisations of parton
distributions.  Additionally, the low moments that have previously
been computed directly from local operators can be used as input into
our approach.  In order to extract these higher moments, we must be
careful of higher-twist contributions, as discussed in
Sec.~\ref{sec:heavy}. Since by using a heavy quark we have removed
many higher-twist terms, we might expect that those that remain will be small.
This would be indicated by the fitted value of $\beta$ being small.
For simplicity, we work to zeroth order in the QCD coupling, taking
the Wilson coefficients to be unity. In a complete analysis, the
perturbative Wilson coefficients appropriate to the desired scheme and
scale
\cite{Gottschalk:1980rv,Kramer:1991uh,Aivazis:1993kh,Aivazis:1993pi}
should be used. With this assumption, the only unknowns in
Eq.~(\ref{eq:12}) are the moments $A_{\psi}^{n} (\mu)$ and the
parameters $\alpha$ and $\beta$.  Then if we wish to determine six
moments ($n=2,4,\ldots,12$) for example, we need the Compton amplitude
evaluated at eight or more different combinations of momenta and heavy
quark masses.  {\it A priori}, we would also want to choose the masses
and momenta such that the convergence of the expansion on the RHS of
Eq.~(\ref{eq:12}) allows us to neglect terms beyond a certain
$n_{max}$ and extract the moments for $n<n_{max}$. For example,
choosing $M_{\Psi}=3.54$~GeV, $q_{0}=2.76$~GeV and $Q^2=1.5$~GeV$^2$,
the expansion on the RHS of Eq.~(\ref{eq:12}) falls off as shown in
the left panel of Fig.~\ref{fig:fn}.  However, from experimental
measurements and perturbative counting rules, we know that the moments
of the parton distributions in fact fall off rapidly as $n$ increases
and it may be more useful to choose the masses and momenta so that
$f(n)$ decreases slowly, allowing the natural suppression of the
moments to control how many can be extracted. In this case, choosing
$M_{\Psi}=2.1$~GeV, $q_{0}=1.98$~GeV and $Q^2=-3.85$~GeV$^2$ which
gives the flatter behaviour shown in the RHS of Fig.~\ref{fig:fn} may
be more appropriate. Without performing the large scale simulations
that are required to determine the Compton amplitude, it is hard to be
definite on the choices of parameters. However, it seems that this
approach has significant potential to determine higher moments of
isovector parton distributions than are currently available from QCD.


\section{Distribution amplitudes from current-current matrix elements}
\label{sec:DAs}

A further application of the approach we have outlined is in computing
moments of meson distribution amplitudes, $\phi_M$. In the lattice
approach, we can extract moments of distribution amplitudes in the
same way as DIS determines moments of parton distributions; for
example, we may study the matrix element $\langle \pi^\pm|T[
V^\mu_{\Psi,\psi}(x) A^\nu_{\Psi,\psi}(0)]|0\rangle$, where
$V^\mu_{\Psi,\psi}$ and $A^\mu_{\Psi,\psi}$ are fictitious vector and
axial vector heavy-light currents.  This process is described by the
tensor
\begin{eqnarray}
    \label{eq:18}
    S^{\mu\nu}_{\Psi,\psi}(p,q)=\int d^4x\, e^{i\,q\cdot x}
    \langle\pi^+(p)|T [V^{\mu}_{\Psi,\psi}(x)
  A^\nu_{\Psi,\psi}(0)]|0\rangle\,.
\end{eqnarray}
Following from Eq.(\ref{eq:5}), the OPE of the two currents leads to
the same matrix elements of twist-two operators that determine the
moments of the pion distribution amplitude:
\begin{eqnarray}
  \label{eq:17}
  \langle \pi^+(p)|\overline\psi \gamma^{\{\mu_1}\gamma_5
  (i\,D)^{\mu_2}\ldots (i\,D)^{\mu_n\}}\psi |0\rangle\, &=&
  f_\pi\langle\xi^{n-1}\rangle_\pi  \left[p^{\mu_1}\ldots p^{\mu_n}  -
    {\rm traces}\right] \,,
\end{eqnarray}
where
\begin{eqnarray}
\label{eq:25}
\langle\xi^n\rangle_\pi\equiv \int_{0}^{1}d\xi\, \xi^n \phi_\pi(\xi)\,.
\end{eqnarray}

\begin{figure}[!t]
  \centering \includegraphics[width=0.3\columnwidth]{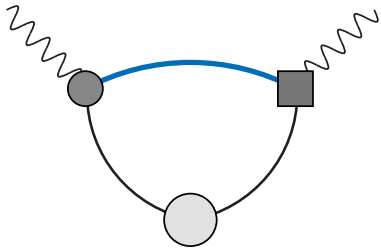}
  \caption{Extraction of moments of meson distribution
    amplitudes. Here, the light-shaded circle denotes the pion
    interpolating operator and the dark circle and dark square
    indicate the vector and axial-vector currents, respectively.  }
  \label{fig:DA}
\end{figure}

These matrix elements can be determined by studying the various
components of $S^{\mu\nu}_{\Psi,\psi}$ for varying $m_\Psi$ and
$q^\mu$ analogously to Eq.~(\ref{eq:9}).  As in the DIS case, many
higher-twist contributions are absent because of the valence nature
heavy quark and the problems that plague direct evaluation of higher
moments due to the lattice cutoff are eliminated.  Since only the
zeroth (decay constant) and second moments of the pion distribution
amplitude have been investigated in the direct approach
\cite{Gottlieb:1986ie,Martinelli:1987si,DeGrand:1987vy,Daniel:1990ah,
  DelDebbio:1999mq,DelDebbio:2002mq}, any information on higher
moments will be useful in constraining the distribution amplitude from
QCD.  For flavour non-diagonal mesons ({\it e.g.} $\pi^\pm$,
$K^{\pm,0}$), extraction of the tensor $S^{\mu\nu}_{\Psi,\psi}$ on the
lattice only requires the computation of the Wick contraction shown in
Fig.~\ref{fig:DA}.


\section{Summary}
\label{sec:summary}

To summarise, the direct study of Compton scattering tensor on the
lattice using the operator product expansion can provide useful
information on the moments of quark distributions.  Using currents
that couple an unphysical, quenched, heavy quark field to the physical
light quarks renders the approach feasible without modifying the
non-perturbative physics that can be extracted.  This has the
potential that a large enough number of moments can be extracted that
the parton distributions can be reliably reconstructed from lattice
calculations. Our analysis has focused on the unpolarised isovector
quark distribution, but it can also be used to study the other
twist-two and twist-three parton distributions and generalised parton
distributions.  Additionally, this method will allow computations of
the moments of meson distribution amplitudes where even the lowest
non-trivial moment is not known reliably from the lattice.

\acknowledgements{We are grateful to W.~Melnitchouk, G.~A.~Miller,
  D.~B.~Renner, S.~R.~Sharpe, J.~I.~Skullerud and R.~D.~Young for
  useful discussions. CJDL acknowledges the support of the National
  Centre for Theoretical Sciences, Taipei. This work was supported by
  DOE grants DE-FG03-97ER41014, DE-FG03-00ER41132 and
  DE-FG03-96ER40956.

  
  \bibliography{ope_arxiv_2}

\begin{thebibliography}{56}
\expandafter\ifx\csname natexlab\endcsname\relax\def\natexlab#1{#1}\fi
\expandafter\ifx\csname bibnamefont\endcsname\relax
  \def\bibnamefont#1{#1}\fi
\expandafter\ifx\csname bibfnamefont\endcsname\relax
  \def\bibfnamefont#1{#1}\fi
\expandafter\ifx\csname citenamefont\endcsname\relax
  \def\citenamefont#1{#1}\fi
\expandafter\ifx\csname url\endcsname\relax
  \def\url#1{\texttt{#1}}\fi
\expandafter\ifx\csname urlprefix\endcsname\relax\def\urlprefix{URL }\fi
\providecommand{\bibinfo}[2]{#2}
\providecommand{\eprint}[2][]{\url{#2}}

\bibitem[{\citenamefont{{U.~Aglietti {\it et al.}}}(1998)}]{Aglietti:1998mz}
\bibinfo{author}{\bibnamefont{{U.~Aglietti {\it et al.}}}},
  \bibinfo{journal}{Phys. Lett.} \textbf{\bibinfo{volume}{B432}},
  \bibinfo{pages}{411} (\bibinfo{year}{1998}), \eprint{hep-ph/9804416}.

\bibitem[{\citenamefont{Liu}(2000)}]{Phys.Rev.D62.074501}
\bibinfo{author}{\bibfnamefont{K.-F.} \bibnamefont{Liu}},
  \bibinfo{journal}{Phys. Rev.} \textbf{\bibinfo{volume}{D62}},
  \bibinfo{pages}{074501} (\bibinfo{year}{2000}), \eprint{hep-ph/9910306}.

\bibitem[{\citenamefont{Kronfeld and Photiadis}(1985)}]{Kronfeld:1984zv}
\bibinfo{author}{\bibfnamefont{A.~S.} \bibnamefont{Kronfeld}} \bibnamefont{and}
  \bibinfo{author}{\bibfnamefont{D.~M.} \bibnamefont{Photiadis}},
  \bibinfo{journal}{Phys. Rev.} \textbf{\bibinfo{volume}{D31}},
  \bibinfo{pages}{2939} (\bibinfo{year}{1985}).

\bibitem[{\citenamefont{Martinelli and
  Sachrajda}(1987{\natexlab{a}})}]{Martinelli:1987zd}
\bibinfo{author}{\bibfnamefont{G.}~\bibnamefont{Martinelli}} \bibnamefont{and}
  \bibinfo{author}{\bibfnamefont{C.~T.} \bibnamefont{Sachrajda}},
  \bibinfo{journal}{Phys. Lett.} \textbf{\bibinfo{volume}{B196}},
  \bibinfo{pages}{184} (\bibinfo{year}{1987}{\natexlab{a}}).

\bibitem[{\citenamefont{Wilson}(1969)}]{Phys.Rev.179.1499}
\bibinfo{author}{\bibfnamefont{K.~G.} \bibnamefont{Wilson}},
  \bibinfo{journal}{Phys. Rev.} \textbf{\bibinfo{volume}{179}},
  \bibinfo{pages}{1499} (\bibinfo{year}{1969}).

\bibitem[{\citenamefont{Christ et~al.}(1972)\citenamefont{Christ, Hasslacher,
  and Mueller}}]{Christ:1972ms}
\bibinfo{author}{\bibfnamefont{N.~H.} \bibnamefont{Christ}},
  \bibinfo{author}{\bibfnamefont{B.}~\bibnamefont{Hasslacher}},
  \bibnamefont{and} \bibinfo{author}{\bibfnamefont{A.~H.}
  \bibnamefont{Mueller}}, \bibinfo{journal}{Phys. Rev.}
  \textbf{\bibinfo{volume}{D6}}, \bibinfo{pages}{3543} (\bibinfo{year}{1972}).

\bibitem[{\citenamefont{Gross and Wilczek}(1973)}]{Gross:1973ju}
\bibinfo{author}{\bibfnamefont{D.~J.} \bibnamefont{Gross}} \bibnamefont{and}
  \bibinfo{author}{\bibfnamefont{F.}~\bibnamefont{Wilczek}},
  \bibinfo{journal}{Phys. Rev.} \textbf{\bibinfo{volume}{D8}},
  \bibinfo{pages}{3633} (\bibinfo{year}{1973}).

\bibitem[{\citenamefont{Gross and Wilczek}(1974)}]{Gross:1974cs}
\bibinfo{author}{\bibfnamefont{D.~J.} \bibnamefont{Gross}} \bibnamefont{and}
  \bibinfo{author}{\bibfnamefont{F.}~\bibnamefont{Wilczek}},
  \bibinfo{journal}{Phys. Rev.} \textbf{\bibinfo{volume}{D9}},
  \bibinfo{pages}{980} (\bibinfo{year}{1974}).

\bibitem[{\citenamefont{Georgi and Politzer}(1974)}]{Georgi:1951sr}
\bibinfo{author}{\bibfnamefont{H.}~\bibnamefont{Georgi}} \bibnamefont{and}
  \bibinfo{author}{\bibfnamefont{H.~D.} \bibnamefont{Politzer}},
  \bibinfo{journal}{Phys. Rev.} \textbf{\bibinfo{volume}{D9}},
  \bibinfo{pages}{416} (\bibinfo{year}{1974}).

\bibitem[{\citenamefont{{M.~G\"{o}ckeler {\it et
  al.}}}(2001)}]{Gockeler:2000ja}
\bibinfo{author}{\bibnamefont{{M.~G\"{o}ckeler {\it et al.}}}},
  \bibinfo{journal}{Phys. Rev.} \textbf{\bibinfo{volume}{D63}},
  \bibinfo{pages}{074506} (\bibinfo{year}{2001}), \eprint{hep-lat/0011091}.

\bibitem[{\citenamefont{{M.~G\"{o}ckeler {\it et
  al.}}}(2005)}]{Gockeler:2005vw}
\bibinfo{author}{\bibnamefont{{M.~G\"{o}ckeler {\it et al.}}}}
  (\bibinfo{year}{2005}), \eprint{hep-lat/0506017}.

\bibitem[{\citenamefont{Gockeler et~al.}(2002)}]{Gockeler:2001xw}
\bibinfo{author}{\bibfnamefont{M.}~\bibnamefont{Gockeler}}
  \bibnamefont{et~al.}, \bibinfo{journal}{Nucl. Phys.}
  \textbf{\bibinfo{volume}{B623}}, \bibinfo{pages}{287} (\bibinfo{year}{2002}),
  \eprint{hep-lat/0103038}.

\bibitem[{\citenamefont{Martinelli et~al.}(1995)\citenamefont{Martinelli,
  Pittori, Sachrajda, Testa, and Vladikas}}]{Martinelli:1994ty}
\bibinfo{author}{\bibfnamefont{G.}~\bibnamefont{Martinelli}},
  \bibinfo{author}{\bibfnamefont{C.}~\bibnamefont{Pittori}},
  \bibinfo{author}{\bibfnamefont{C.~T.} \bibnamefont{Sachrajda}},
  \bibinfo{author}{\bibfnamefont{M.}~\bibnamefont{Testa}}, \bibnamefont{and}
  \bibinfo{author}{\bibfnamefont{A.}~\bibnamefont{Vladikas}},
  \bibinfo{journal}{Nucl. Phys.} \textbf{\bibinfo{volume}{B445}},
  \bibinfo{pages}{81} (\bibinfo{year}{1995}), \eprint{hep-lat/9411010}.

\bibitem[{\citenamefont{{M.~G\"{o}ckeler {\it et
  al.}}}(1999)}]{Gockeler:1998ye}
\bibinfo{author}{\bibnamefont{{M.~G\"{o}ckeler {\it et al.}}}},
  \bibinfo{journal}{Nucl. Phys.} \textbf{\bibinfo{volume}{B544}},
  \bibinfo{pages}{699} (\bibinfo{year}{1999}), \eprint{hep-lat/9807044}.

\bibitem[{\citenamefont{Capitani}(2003)}]{Capitani:2002mp}
\bibinfo{author}{\bibfnamefont{S.}~\bibnamefont{Capitani}},
  \bibinfo{journal}{Phys. Rept.} \textbf{\bibinfo{volume}{382}},
  \bibinfo{pages}{113} (\bibinfo{year}{2003}), \eprint{hep-lat/0211036}.

\bibitem[{\citenamefont{{M.~Guagnelli {\it et al.}}}(2005)}]{Guagnelli:2004ga}
\bibinfo{author}{\bibnamefont{{M.~Guagnelli {\it et al.}}}}
  (\bibinfo{collaboration}{Zeuthen-Rome (ZeRo)}), \bibinfo{journal}{Eur. Phys.
  J.} \textbf{\bibinfo{volume}{C40}}, \bibinfo{pages}{69}
  (\bibinfo{year}{2005}), \eprint{hep-lat/0405027}.

\bibitem[{\citenamefont{{C.~Dawson {\it et al.}}}(1998)}]{Dawson:1997ic}
\bibinfo{author}{\bibnamefont{{C.~Dawson {\it et al.}}}},
  \bibinfo{journal}{Nucl. Phys.} \textbf{\bibinfo{volume}{B514}},
  \bibinfo{pages}{313} (\bibinfo{year}{1998}), \eprint{hep-lat/9707009}.

\bibitem[{\citenamefont{{S.~Capitani {\it et
  al.}}}(1999{\natexlab{a}})}]{Capitani:1998fe}
\bibinfo{author}{\bibnamefont{{S.~Capitani {\it et al.}}}},
  \bibinfo{journal}{Nucl. Phys. Proc. Suppl.} \textbf{\bibinfo{volume}{73}},
  \bibinfo{pages}{288} (\bibinfo{year}{1999}{\natexlab{a}}),
  \eprint{hep-lat/9809171}.

\bibitem[{\citenamefont{{S.~Capitani {\it et
  al.}}}(1999{\natexlab{b}})}]{Capitani:1999fm}
\bibinfo{author}{\bibnamefont{{S.~Capitani {\it et al.}}}},
  \bibinfo{journal}{Nucl. Phys. Proc. Suppl.} \textbf{\bibinfo{volume}{79}},
  \bibinfo{pages}{173} (\bibinfo{year}{1999}{\natexlab{b}}),
  \eprint{hep-ph/9906320}.

\bibitem[{\citenamefont{Shifman}(1994)}]{Shifman:1994yf}
\bibinfo{author}{\bibfnamefont{M.~A.} \bibnamefont{Shifman}}
  (\bibinfo{year}{1994}), \eprint{hep-ph/9405246}.

\bibitem[{\citenamefont{Detmold
  et~al.}(2001{\natexlab{a}})\citenamefont{Detmold, Melnitchouk, Negele,
  Renner, and Thomas}}]{Detmold:2001jb}
\bibinfo{author}{\bibfnamefont{W.}~\bibnamefont{Detmold}},
  \bibinfo{author}{\bibfnamefont{W.}~\bibnamefont{Melnitchouk}},
  \bibinfo{author}{\bibfnamefont{J.~W.} \bibnamefont{Negele}},
  \bibinfo{author}{\bibfnamefont{D.~B.} \bibnamefont{Renner}},
  \bibnamefont{and} \bibinfo{author}{\bibfnamefont{A.~W.}
  \bibnamefont{Thomas}}, \bibinfo{journal}{Phys. Rev. Lett.}
  \textbf{\bibinfo{volume}{87}}, \bibinfo{pages}{172001}
  (\bibinfo{year}{2001}{\natexlab{a}}), \eprint{hep-lat/0103006}.

\bibitem[{\citenamefont{Arndt and Savage}(2002)}]{Arndt:2001ye}
\bibinfo{author}{\bibfnamefont{D.}~\bibnamefont{Arndt}} \bibnamefont{and}
  \bibinfo{author}{\bibfnamefont{M.~J.} \bibnamefont{Savage}},
  \bibinfo{journal}{Nucl. Phys.} \textbf{\bibinfo{volume}{A697}},
  \bibinfo{pages}{429} (\bibinfo{year}{2002}), \eprint{nucl-th/0105045}.

\bibitem[{\citenamefont{Chen and Ji}(2001)}]{Chen:2001eg}
\bibinfo{author}{\bibfnamefont{J.-W.} \bibnamefont{Chen}} \bibnamefont{and}
  \bibinfo{author}{\bibfnamefont{X.-D.} \bibnamefont{Ji}},
  \bibinfo{journal}{Phys. Lett.} \textbf{\bibinfo{volume}{B523}},
  \bibinfo{pages}{107} (\bibinfo{year}{2001}), \eprint{hep-ph/0105197}.

\bibitem[{\citenamefont{Chen and Savage}(2002{\natexlab{a}})}]{Chen:2001gr}
\bibinfo{author}{\bibfnamefont{J.-W.} \bibnamefont{Chen}} \bibnamefont{and}
  \bibinfo{author}{\bibfnamefont{M.~J.} \bibnamefont{Savage}},
  \bibinfo{journal}{Nucl. Phys.} \textbf{\bibinfo{volume}{A707}},
  \bibinfo{pages}{452} (\bibinfo{year}{2002}{\natexlab{a}}),
  \eprint{nucl-th/0108042}.

\bibitem[{\citenamefont{Beane and Savage}(2002)}]{Beane:2002vq}
\bibinfo{author}{\bibfnamefont{S.~R.} \bibnamefont{Beane}} \bibnamefont{and}
  \bibinfo{author}{\bibfnamefont{M.~J.} \bibnamefont{Savage}},
  \bibinfo{journal}{Nucl. Phys.} \textbf{\bibinfo{volume}{A709}},
  \bibinfo{pages}{319} (\bibinfo{year}{2002}), \eprint{hep-lat/0203003}.

\bibitem[{\citenamefont{Chen and Savage}(2002{\natexlab{b}})}]{Chen:2001yi}
\bibinfo{author}{\bibfnamefont{J.-W.} \bibnamefont{Chen}} \bibnamefont{and}
  \bibinfo{author}{\bibfnamefont{M.~J.} \bibnamefont{Savage}},
  \bibinfo{journal}{Phys. Rev.} \textbf{\bibinfo{volume}{D65}},
  \bibinfo{pages}{094001} (\bibinfo{year}{2002}{\natexlab{b}}),
  \eprint{hep-lat/0111050}.

\bibitem[{\citenamefont{Detmold et~al.}(2002)\citenamefont{Detmold,
  Melnitchouk, and Thomas}}]{Detmold:2002nf}
\bibinfo{author}{\bibfnamefont{W.}~\bibnamefont{Detmold}},
  \bibinfo{author}{\bibfnamefont{W.}~\bibnamefont{Melnitchouk}},
  \bibnamefont{and} \bibinfo{author}{\bibfnamefont{A.~W.}
  \bibnamefont{Thomas}}, \bibinfo{journal}{Phys. Rev.}
  \textbf{\bibinfo{volume}{D66}}, \bibinfo{pages}{054501}
  (\bibinfo{year}{2002}), \eprint{hep-lat/0206001}.

\bibitem[{\citenamefont{Detmold et~al.}(2003)\citenamefont{Detmold,
  Melnitchouk, and Thomas}}]{Detmold:2003tm}
\bibinfo{author}{\bibfnamefont{W.}~\bibnamefont{Detmold}},
  \bibinfo{author}{\bibfnamefont{W.}~\bibnamefont{Melnitchouk}},
  \bibnamefont{and} \bibinfo{author}{\bibfnamefont{A.~W.}
  \bibnamefont{Thomas}}, \bibinfo{journal}{Phys. Rev.}
  \textbf{\bibinfo{volume}{D68}}, \bibinfo{pages}{034025}
  (\bibinfo{year}{2003}), \eprint{hep-lat/0303015}.

\bibitem[{\citenamefont{Detmold and Lin}(2005)}]{Detmold:2005pt}
\bibinfo{author}{\bibfnamefont{W.}~\bibnamefont{Detmold}} \bibnamefont{and}
  \bibinfo{author}{\bibfnamefont{C.~J.~D.} \bibnamefont{Lin}},
  \bibinfo{journal}{Phys. Rev.} \textbf{\bibinfo{volume}{D71}},
  \bibinfo{pages}{054510} (\bibinfo{year}{2005}), \eprint{hep-lat/0501007}.

\bibitem[{\citenamefont{Nachtmann}(1973)}]{Nachtmann:1973mr}
\bibinfo{author}{\bibfnamefont{O.}~\bibnamefont{Nachtmann}},
  \bibinfo{journal}{Nucl. Phys.} \textbf{\bibinfo{volume}{B63}},
  \bibinfo{pages}{237} (\bibinfo{year}{1973}).

\bibitem[{\citenamefont{Bloom and Gilman}(1970)}]{Bloom:1970xb}
\bibinfo{author}{\bibfnamefont{E.~D.} \bibnamefont{Bloom}} \bibnamefont{and}
  \bibinfo{author}{\bibfnamefont{F.~J.} \bibnamefont{Gilman}},
  \bibinfo{journal}{Phys. Rev. Lett.} \textbf{\bibinfo{volume}{25}},
  \bibinfo{pages}{1140} (\bibinfo{year}{1970}).

\bibitem[{\citenamefont{Bloom and Gilman}(1971)}]{Bloom:1971ye}
\bibinfo{author}{\bibfnamefont{E.~D.} \bibnamefont{Bloom}} \bibnamefont{and}
  \bibinfo{author}{\bibfnamefont{F.~J.} \bibnamefont{Gilman}},
  \bibinfo{journal}{Phys. Rev.} \textbf{\bibinfo{volume}{D4}},
  \bibinfo{pages}{2901} (\bibinfo{year}{1971}).

\bibitem[{\citenamefont{Georgi and
  Politzer}(1976{\natexlab{a}})}]{Georgi:1976vf}
\bibinfo{author}{\bibfnamefont{H.}~\bibnamefont{Georgi}} \bibnamefont{and}
  \bibinfo{author}{\bibfnamefont{H.~D.} \bibnamefont{Politzer}},
  \bibinfo{journal}{Phys. Rev. Lett.} \textbf{\bibinfo{volume}{36}},
  \bibinfo{pages}{1281} (\bibinfo{year}{1976}{\natexlab{a}}).

\bibitem[{\citenamefont{Georgi and
  Politzer}(1976{\natexlab{b}})}]{Georgi:1976ve}
\bibinfo{author}{\bibfnamefont{H.}~\bibnamefont{Georgi}} \bibnamefont{and}
  \bibinfo{author}{\bibfnamefont{H.~D.} \bibnamefont{Politzer}},
  \bibinfo{journal}{Phys. Rev.} \textbf{\bibinfo{volume}{D14}},
  \bibinfo{pages}{1829} (\bibinfo{year}{1976}{\natexlab{b}}).

\bibitem[{\citenamefont{Wandzura}(1977)}]{Wandzura:1977ce}
\bibinfo{author}{\bibfnamefont{S.}~\bibnamefont{Wandzura}},
  \bibinfo{journal}{Nucl. Phys.} \textbf{\bibinfo{volume}{B122}},
  \bibinfo{pages}{412} (\bibinfo{year}{1977}).

\bibitem[{\citenamefont{Witten}(1976)}]{Witten:1975bh}
\bibinfo{author}{\bibfnamefont{E.}~\bibnamefont{Witten}},
  \bibinfo{journal}{Nucl. Phys.} \textbf{\bibinfo{volume}{B104}},
  \bibinfo{pages}{445} (\bibinfo{year}{1976}).

\bibitem[{\citenamefont{El-Khadra and Luke}(2002)}]{El-Khadra:2002wp}
\bibinfo{author}{\bibfnamefont{A.~X.} \bibnamefont{El-Khadra}}
  \bibnamefont{and} \bibinfo{author}{\bibfnamefont{M.}~\bibnamefont{Luke}},
  \bibinfo{journal}{Ann. Rev. Nucl. Part. Sci.} \textbf{\bibinfo{volume}{52}},
  \bibinfo{pages}{201} (\bibinfo{year}{2002}), \eprint{hep-ph/0208114}.

\bibitem[{\citenamefont{Gottschalk}(1981)}]{Gottschalk:1980rv}
\bibinfo{author}{\bibfnamefont{T.}~\bibnamefont{Gottschalk}},
  \bibinfo{journal}{Phys. Rev.} \textbf{\bibinfo{volume}{D23}},
  \bibinfo{pages}{56} (\bibinfo{year}{1981}).

\bibitem[{\citenamefont{Kramer and Lampe}(1992)}]{Kramer:1991uh}
\bibinfo{author}{\bibfnamefont{G.}~\bibnamefont{Kramer}} \bibnamefont{and}
  \bibinfo{author}{\bibfnamefont{B.}~\bibnamefont{Lampe}}, \bibinfo{journal}{Z.
  Phys.} \textbf{\bibinfo{volume}{C54}}, \bibinfo{pages}{139}
  (\bibinfo{year}{1992}).

\bibitem[{\citenamefont{Aivazis
  et~al.}(1994{\natexlab{a}})\citenamefont{Aivazis, Olness, and
  Tung}}]{Aivazis:1993kh}
\bibinfo{author}{\bibfnamefont{M.~A.~G.} \bibnamefont{Aivazis}},
  \bibinfo{author}{\bibfnamefont{F.~I.} \bibnamefont{Olness}},
  \bibnamefont{and} \bibinfo{author}{\bibfnamefont{W.-K.} \bibnamefont{Tung}},
  \bibinfo{journal}{Phys. Rev.} \textbf{\bibinfo{volume}{D50}},
  \bibinfo{pages}{3085} (\bibinfo{year}{1994}{\natexlab{a}}),
  \eprint{hep-ph/9312318}.

\bibitem[{\citenamefont{Aivazis
  et~al.}(1994{\natexlab{b}})\citenamefont{Aivazis, Collins, Olness, and
  Tung}}]{Aivazis:1993pi}
\bibinfo{author}{\bibfnamefont{M.~A.~G.} \bibnamefont{Aivazis}},
  \bibinfo{author}{\bibfnamefont{J.~C.} \bibnamefont{Collins}},
  \bibinfo{author}{\bibfnamefont{F.~I.} \bibnamefont{Olness}},
  \bibnamefont{and} \bibinfo{author}{\bibfnamefont{W.-K.} \bibnamefont{Tung}},
  \bibinfo{journal}{Phys. Rev.} \textbf{\bibinfo{volume}{D50}},
  \bibinfo{pages}{3102} (\bibinfo{year}{1994}{\natexlab{b}}),
  \eprint{hep-ph/9312319}.

\bibitem[{\citenamefont{Jaffe and Ji}(1991)}]{Jaffe:1991kp}
\bibinfo{author}{\bibfnamefont{R.~L.} \bibnamefont{Jaffe}} \bibnamefont{and}
  \bibinfo{author}{\bibfnamefont{X.-D.} \bibnamefont{Ji}},
  \bibinfo{journal}{Phys. Rev. Lett.} \textbf{\bibinfo{volume}{67}},
  \bibinfo{pages}{552} (\bibinfo{year}{1991}).

\bibitem[{\citenamefont{Jaffe and Ji}(1992)}]{Jaffe:1991ra}
\bibinfo{author}{\bibfnamefont{R.~L.} \bibnamefont{Jaffe}} \bibnamefont{and}
  \bibinfo{author}{\bibfnamefont{X.-D.} \bibnamefont{Ji}},
  \bibinfo{journal}{Nucl. Phys.} \textbf{\bibinfo{volume}{B375}},
  \bibinfo{pages}{527} (\bibinfo{year}{1992}).

\bibitem[{\citenamefont{{H.~Avakian {\it et al.}}}(2004)}]{Avakian:2003pk}
\bibinfo{author}{\bibnamefont{{H.~Avakian {\it et al.}}}}
  (\bibinfo{collaboration}{CLAS}), \bibinfo{journal}{Phys. Rev.}
  \textbf{\bibinfo{volume}{D69}}, \bibinfo{pages}{112004}
  (\bibinfo{year}{2004}), \eprint{hep-ex/0301005}.

\bibitem[{\citenamefont{Efremov et~al.}(2003)\citenamefont{Efremov, Goeke, and
  Schweitzer}}]{Efremov:2002ut}
\bibinfo{author}{\bibfnamefont{A.~V.} \bibnamefont{Efremov}},
  \bibinfo{author}{\bibfnamefont{K.}~\bibnamefont{Goeke}}, \bibnamefont{and}
  \bibinfo{author}{\bibfnamefont{P.}~\bibnamefont{Schweitzer}},
  \bibinfo{journal}{Phys. Rev.} \textbf{\bibinfo{volume}{D67}},
  \bibinfo{pages}{114014} (\bibinfo{year}{2003}), \eprint{hep-ph/0208124}.

\bibitem[{\citenamefont{{J.~Bl\"{u}mlein}}(2001)}]{Blumlein:2001ca}
\bibinfo{author}{\bibnamefont{{J.~Bl\"{u}mlein}}}, \bibinfo{journal}{Eur. Phys.
  J.} \textbf{\bibinfo{volume}{C20}}, \bibinfo{pages}{683}
  (\bibinfo{year}{2001}), \eprint{hep-ph/0104099}.

\bibitem[{\citenamefont{Ioffe and Khodjamirian}(1995)}]{Ioffe:1994aa}
\bibinfo{author}{\bibfnamefont{B.~L.} \bibnamefont{Ioffe}} \bibnamefont{and}
  \bibinfo{author}{\bibfnamefont{A.}~\bibnamefont{Khodjamirian}},
  \bibinfo{journal}{Phys. Rev.} \textbf{\bibinfo{volume}{D51}},
  \bibinfo{pages}{3373} (\bibinfo{year}{1995}), \eprint{hep-ph/9403371}.

\bibitem[{\citenamefont{Liu and Dong}(1994)}]{Liu:1993cv}
\bibinfo{author}{\bibfnamefont{K.-F.} \bibnamefont{Liu}} \bibnamefont{and}
  \bibinfo{author}{\bibfnamefont{S.-J.} \bibnamefont{Dong}},
  \bibinfo{journal}{Phys. Rev. Lett.} \textbf{\bibinfo{volume}{72}},
  \bibinfo{pages}{1790} (\bibinfo{year}{1994}), \eprint{hep-ph/9306299}.

\bibitem[{\citenamefont{Titchmarsh}(1960)}]{CarlsonThm}
\bibinfo{author}{\bibfnamefont{E.~C.} \bibnamefont{Titchmarsh}},
  \emph{\bibinfo{title}{The theory of functions}} (\bibinfo{publisher}{Oxford
  University Press}, \bibinfo{year}{1960}), \bibinfo{edition}{2nd} ed.

\bibitem[{\citenamefont{Detmold
  et~al.}(2001{\natexlab{b}})\citenamefont{Detmold, Melnitchouk, and
  Thomas}}]{Detmold:2001dv}
\bibinfo{author}{\bibfnamefont{W.}~\bibnamefont{Detmold}},
  \bibinfo{author}{\bibfnamefont{W.}~\bibnamefont{Melnitchouk}},
  \bibnamefont{and} \bibinfo{author}{\bibfnamefont{A.~W.}
  \bibnamefont{Thomas}}, \bibinfo{journal}{Eur. Phys. J. direct}
  \textbf{\bibinfo{volume}{C3}}, \bibinfo{pages}{13}
  (\bibinfo{year}{2001}{\natexlab{b}}), \eprint{hep-lat/0108002}.

\bibitem[{\citenamefont{Gottlieb and Kronfeld}(1986)}]{Gottlieb:1986ie}
\bibinfo{author}{\bibfnamefont{S.~A.} \bibnamefont{Gottlieb}} \bibnamefont{and}
  \bibinfo{author}{\bibfnamefont{A.~S.} \bibnamefont{Kronfeld}},
  \bibinfo{journal}{Phys. Rev.} \textbf{\bibinfo{volume}{D33}},
  \bibinfo{pages}{227} (\bibinfo{year}{1986}).

\bibitem[{\citenamefont{Martinelli and
  Sachrajda}(1987{\natexlab{b}})}]{Martinelli:1987si}
\bibinfo{author}{\bibfnamefont{G.}~\bibnamefont{Martinelli}} \bibnamefont{and}
  \bibinfo{author}{\bibfnamefont{C.~T.} \bibnamefont{Sachrajda}},
  \bibinfo{journal}{Phys. Lett.} \textbf{\bibinfo{volume}{B190}},
  \bibinfo{pages}{151} (\bibinfo{year}{1987}{\natexlab{b}}).

\bibitem[{\citenamefont{DeGrand and Loft}(1988)}]{DeGrand:1987vy}
\bibinfo{author}{\bibfnamefont{T.~A.} \bibnamefont{DeGrand}} \bibnamefont{and}
  \bibinfo{author}{\bibfnamefont{R.~D.} \bibnamefont{Loft}},
  \bibinfo{journal}{Phys. Rev.} \textbf{\bibinfo{volume}{D38}},
  \bibinfo{pages}{954} (\bibinfo{year}{1988}).

\bibitem[{\citenamefont{Daniel et~al.}(1991)\citenamefont{Daniel, Gupta, and
  Richards}}]{Daniel:1990ah}
\bibinfo{author}{\bibfnamefont{D.}~\bibnamefont{Daniel}},
  \bibinfo{author}{\bibfnamefont{R.}~\bibnamefont{Gupta}}, \bibnamefont{and}
  \bibinfo{author}{\bibfnamefont{D.~G.} \bibnamefont{Richards}},
  \bibinfo{journal}{Phys. Rev.} \textbf{\bibinfo{volume}{D43}},
  \bibinfo{pages}{3715} (\bibinfo{year}{1991}).

\bibitem[{\citenamefont{Del~Debbio et~al.}(2000)\citenamefont{Del~Debbio,
  Di~Pierro, Dougall, and Sachrajda}}]{DelDebbio:1999mq}
\bibinfo{author}{\bibfnamefont{L.}~\bibnamefont{Del~Debbio}},
  \bibinfo{author}{\bibfnamefont{M.}~\bibnamefont{Di~Pierro}},
  \bibinfo{author}{\bibfnamefont{A.}~\bibnamefont{Dougall}}, \bibnamefont{and}
  \bibinfo{author}{\bibfnamefont{C.~T.} \bibnamefont{Sachrajda}}
  (\bibinfo{collaboration}{UKQCD}), \bibinfo{journal}{Nucl. Phys. Proc. Suppl.}
  \textbf{\bibinfo{volume}{83}}, \bibinfo{pages}{235} (\bibinfo{year}{2000}),
  \eprint{hep-lat/9909147}.

\bibitem[{\citenamefont{Del~Debbio et~al.}(2003)\citenamefont{Del~Debbio,
  Di~Pierro, and Dougall}}]{DelDebbio:2002mq}
\bibinfo{author}{\bibfnamefont{L.}~\bibnamefont{Del~Debbio}},
  \bibinfo{author}{\bibfnamefont{M.}~\bibnamefont{Di~Pierro}},
  \bibnamefont{and} \bibinfo{author}{\bibfnamefont{A.}~\bibnamefont{Dougall}},
  \bibinfo{journal}{Nucl. Phys. Proc. Suppl.} \textbf{\bibinfo{volume}{119}},
  \bibinfo{pages}{416} (\bibinfo{year}{2003}), \eprint{hep-lat/0211037}.

\end{thebibliography}

\end{document}